\documentclass[a4paper,english,elsevier, preprint, superscriptaddress]{revtex4-1}
\usepackage[T1]{fontenc}
\usepackage[latin9]{inputenc}
\usepackage{array}
\usepackage{float}
\usepackage{textcomp}
\usepackage{amsmath}
\usepackage{amssymb}
\usepackage{graphicx}

\makeatletter

\pdfpageheight\paperheight
\pdfpagewidth\paperwidth

\providecommand{\tabularnewline}{\\}

\makeatother

\usepackage{babel}
\begin{document}
\preprint{}
\title{A FRACTAL VIEWPOINT TO COVID-19 INFECTION}
\author{Oscar Sotolongo-Costa}
\email{osotolongo@gmail.com}

\affiliation{C\'atedra \textquotedbl Henri Poincar\'e\textquotedbl{} de sistemas complejos. Universidad
de La Habana, Habana 10400 Cuba.}
\author{Jos\'e Weberszpil}
\email{josewebe@gmail.com}

\affiliation{Universidade Federal Rural do Rio de Janeiro, UFRRJ-IM/DTL;~ Av. Governador Roberto Silveira s/n- Nova Igua\c{c}\'u, Rio de Janeiro,
Brasil 695014.}
\author{Oscar Sotolongo-Grau}
\email{osotolongo@fundacioace.com}

\affiliation{Alzheimer Research Center and Memory Clinic, Fundaci\'0 ACE, Institut
Catal\'a de Neuroci\'encies Aplicades;~\\
 08029 Barcelona, Spain.}
\date{\today}
\begin{abstract}
One of the central tools to control the COVID-19 pandemics is the
knowledge of its spreading dynamics. Here we develop a fractal model
capable of describe this dynamics, in term of daily new cases, and
provide quantitative criteria for some predictions. We propose a fractal
dynamical model using conformed derivative and fractal time scale.
A Burr-XII shaped solution of the fractal-like equation is obtained.
The model is tested using data from several countries, showing that
a single function is able to describe very different shapes of the
outbreak. The diverse behavior of the outbreak on those countries
is presented and discussed. Moreover, a criterion to determine the
existence of the pandemic peak and a expression to find the time to
reach herd immunity are also obtained.
\begin{description}
\item [{PACS~numbers}] 05.90.+m - Other topics in statistical physics,
thermodynamics, and nonlinear dynamical systems, 11.10.Lm - Nonlinear
or nonlocal theories and models, 02.90.+p - Other topics in mathematical
methods in physics.
\item [{Keywords}] Covid-19; Nonlinear Fractal-Like Kinetics; Analogy Relaxation
Model; Nonlinear Equations; Herd Immunity; Deformed Derivatives.
\end{description}
\end{abstract}

\maketitle

\section{Introduction }

The worldwide pandemic provoked by the SARS-CoV-2 coronavirus outbreak
have attracted the attention of the scientific community due to, among
other features, its fast spread. Its strong contamination capacity
has created a fast growing population of people enduring COVID-19,
its related disease, and a non small peak of mortality. The temporal
evolution of contagion over different countries and worldwide brings
up a common dynamic characteristic, in particular, its fast rise to
reach a maximum followed by a slow decrease (incidentally, very similar
to other epidemic processes) suggesting some kind of relaxation process,
which we try to deal with, since relaxation is, essentially, a process
where the parameters characterizing a system are altered, followed
by a tendency to equilibrium values. In Physics, clear examples are,
among others, dielectric or mechanical relaxation. In other fields
(psychology, economy, etc.) there are also phenomena in which an analogy
with \textquotedbl common\textquotedbl{} relaxation can be established.
In relaxation, temporal behavior of parameters is of medular methodological
interest. That is why pandemics can be conceived as one in which this
behavior is also present. For this reason, we are interested, despite
the existence of statistical or dynamical systems method, in the introduction
of a phenomenological equation containing parameters that reflect
the system\textasciiacute s behavior, from which its dynamics emerges.
We are interested in studying the daily presented new cases, not the
current cases by day. This must be noted to avoid confusion in the
interpretation, i.e. we study not the cumulative number of infected
patients reported in databases, but its derivative. This relaxation
process in this case is, for us, an scenario that, by analogy, will
serve to model the dynamics of the pandemics. This is not an ordinary
process. Due to the concurrence of many factors that make very complex
its study, its description must turn out to non classical description.
So, we will consider that the dynamics of this pandemic is described
by a \textquotedbl fractal\textquotedbl{} or internal time \citep{Jap-Weber-Sotolongo}.
The network formed by the people in its daily activity forms a complex
field of links very difficult, if not impossible, to describe. However,
we can take a simplified model where all the nodes belong to a small
world network, but the time of transmission from one node to other
differs for each link. So, in order to study this process let us assume
that spread occurs in \textquotedbl fractal time\textquotedbl{} or
internal time \citep{Jap-Weber-Sotolongo,jonscher1994all}. This is
not a new tool in physics. In refs. \citep{brouers2006generalized,brouers2014fractal,sotolongo2015non}
this concept has been successfully introduced and here, we keep in
mind the possibility of a fractal-like kinetics \citep{schnell2004reaction},
but generalizing as a nonlinear kinetic process. Here we will follow
to what we refer as a \textquotedbl relaxation-like\textquotedbl{}
approach, to model the dynamics of the pandemic and that justify the
fractal time. By analogy with relaxation, an anomalous relaxation,
we build up a simple nonlinear equation with fractal-time. We also
regain the analytical results using a deformed derivative approach,
using conformable derivative (CD) \citep{KHALIL201465}. In Ref. \citep{weberszpil2015connection}
one of the authors (J.W.) have shown intimate relation of this derivative
with complex systems and nonadditive statistical mechanics. This was
done without resort to details of any kind of specific entropy definition. 

Our article is outlined as follows: In Section 2, we present the fractal
model formulated in terms of conformable derivatives, to develop the
relevant expressions to adjust data of COVID-19. In Section 3, we
show the results and figures referring to the data fitting along with
discussions. In section 4, we finally cast our general conclusions
and possible paths for further investigations.

\section{Fractal Model}

Let us denote by $F(t)$ the number of contagions up to time $t$.

The CD is defined as \citep{KHALIL201465} 
\begin{equation}
D_{x}^{\alpha}f(x)=\lim_{\epsilon\rightarrow0}\frac{f(x+\epsilon x^{1-\alpha})-f(x)}{\epsilon}.\label{eq:Deformed Derivative-Deff-1}
\end{equation}

Note that the deformation is placed in the independent variable.

For differentiable functions, the CD can be written as

\begin{equation}
D_{x}^{\alpha}f=x^{1-\alpha}\dfrac{df}{dx}.\label{eq:conformable-deriv-differentiable-1}
\end{equation}

An important point to be noticed here is that the deformations affect
different functional spaces, depending on the problem under consideration.
For the conformable derivative \citep{weberszpil2015connection,weberszpil2016variational,weberszpil2017generalized,weberszpil2017structural,weberszpil2020dual},
the deformations are put in the independent variable, which can be
a space coordinate, in the case of, e.g, mass position dependent problems,
or even time or spacetime variables, for temporal dependent parameter
or relativistic problems. Since we are dealing with a complex system,
a search for a mathematical approach that could take into account
some fractality or hidden variables seems to be adequate. This idea
is also based in the fact that we do not have full information about
the system under study. In this case, deformed derivatives with fractal
time seems to be a good option to deal with this kind of system. Deformed
derivatives, in the context of generalized statistical mechanics are
present and connected \citep{weberszpil2015connection}. There, the
authors have also shown that the $q-deformed$ derivative has also
a dual derivative and a $q-exponential$ related function \citep{rosa2018dual}.
Here, in the case under study, the deformation is considered for the
solutions-space or dependent variable, that is, the number $F(t)$
of contagions up to time $t$. One should also consider that justification
for the use of deformed derivatives finds its physical basis on the
mapping into the fractal continuum \citep{weberszpil2015connection,balankin2012hydrodynamics,balankin2012map,balankin2016towards}.
That is, one considers a mapping from a fractal coarse grained (fractal
porous) space, which is essentially discontinuous in the embedding
Euclidean space, to a continuous one \citep{weberszpil2016variational}.
In our case the fractality lies in the temporal variable. Then the
CD is with respect to time. 

A nonlinear relaxation model can be proposed here, again based on
a generalization of Brouers-Sotolongo fractal kinetic model (BSf)
\citep{brouers2006generalized,brouers2014fractal,brouers2019use},
but here represented by a nonlinear equation written in terms of CD:

\begin{equation}
D_{t}^{\alpha}F=\frac{1}{\tau^{\alpha}}F^{q},\label{eq:NonLinearRelaxationModel}
\end{equation}
 where $\tau$ is our \textquotedbl relaxation time\textquotedbl{}
and $q$ and $\alpha$ here are real parameters. We do not impose
any limit for the parameters. Equation (\ref{eq:NonLinearRelaxationModel})
has as a well known solution a function with the shape of Burr XII
\citep{burr1942cumulative}, with :

\begin{equation}
F=F_{0}[1+(1-q)\frac{(t^{\alpha}-t_{0}^{\alpha})}{\tau^{\alpha}\alpha F_{0}^{1-q}}]^{\frac{1}{1-q}}.\label{eq:Cumulative}
\end{equation}

The density (in a similar form of a PDF, but here it is not a PDF)
is, then:

\begin{equation}
f(t)=\frac{F_{0}^{q}}{\tau^{\alpha}}[C+(1-q)\frac{t^{\alpha}}{\tau^{\alpha}\alpha F_{0}^{1-q}}]^{\frac{q}{1-q}}t^{\alpha-1},
\end{equation}
 where $C=1-\frac{(1-q)t_{0}^{\alpha}}{\tau^{\alpha}\alpha F_{0}^{1-q}},$which
can be expressed as:

\begin{equation}
f(t)=A'[C+B't^{\alpha}]^{-b}t^{a-1}\label{eq:Fitting eq-1}
\end{equation}
 where the parameter are $A'=\frac{F_{0}^{q}}{\tau^{\alpha}},$ $B'=(1-q)\frac{1}{\tau^{\alpha}\alpha F_{0}^{1-q}},$
$b=\frac{q}{q-1},$ $a=\alpha.$

Or, in a simpler form for data adjustment purposes 
\begin{equation}
f(t)=A[1+Bt^{\alpha}]^{-b}t^{a-1},\label{eq:Fitting eq-1-2}
\end{equation}
 with $A=\frac{A'}{C^{b}},$ $B=\frac{B'}{C}.$

This is very similar, though not equal, to the function proposed by
Tsallis \citep{TSALLIS2020.04.24.20078154,tsallis2020predicting}
in an ad hoc way. Here, however, a physical representation by the
method of analogy is proposed to describe the evolution of the pandemics.
Though we have introduced $A,$ $B,$ $C,$ $b$, and $a$ as parameters
to simplify the fitting, the true adjustment constants are, clearly,
$q,$ $\tau$ and $\alpha.$ Note that we do not impose any restrictive
values to the parameters. 

There is no need to demand that the solution always converge. The
equation to obtain Burr XII has to impose restrictions but this is
not the case. In Burr XII the function was used as a probability distribution.
But here the function describes a dynamic, which can be explosive,
as will be shown for the curves of Brazil and Mexico. Therefore, if
we consider infinite population, a peak will never be reached unless
the circumstances change (treatments, vaccines, isolation, etc.).
Our model does not impose finiteness of the solution. The possibility
for a decay of the pandemic in a given region in this model requires
the fulfillment of the condition 
\begin{equation}
a(1-b)-1<0,\label{eq:Condition-1}
\end{equation}
what expresses the property that 
\begin{equation}
\underset{t\rightarrow\infty}{\lim}f(t)=0,\label{eq:Property}
\end{equation}
what means that the function has a local maximum. If this condition
is not accomplished, the pandemic does not have a peak and, therefore,
the number of cases increases forever in this model.

In this case there is, apart from the change of propagation and development
conditions, the possibility for a given country that does not satisfies
condition (\ref{eq:Condition-1}), to reach \textquotedbl herd immunity\textquotedbl ,
i.e., when the number of contagions has reached about 60\% of population,
in which case we may calculate the time to reach such state using
(\ref{eq:Cumulative}), assuming $t_{0}=0$: 
\begin{equation}
T_{hi}=[(0.6P)^{1/(1-b)}-1)/B]^{1/a}.
\end{equation}

We will work with what we will call $T_{1000}$ ahead and that seems
to make more sense and bring more information.

\section{Data fitting}

With eq. (\ref{eq:Fitting eq-1-2}) let us fit the data of the epidemic
worldwide. The data was extracted from Johns Hopkins University \citep{John-Hopkins}
and the website \citep{worldometers} to process the data for several
countries. 

We covered the infected cases taken at Jan 22 as day 1, up to June
13. The behavior of new infected cases by day is shown in figure 1.
The fitting was made with gnuplot 5.2. As it seems, the pandemic shows
some sort of \textquotedbl plateau\textquotedbl , so the present
measures of prevention are not able to eliminate the infection propagation
in a short term, but it can be seen that condition (\ref{eq:Condition-1})
is weakly fulfilled. 
\begin{figure}[H]
\begin{minipage}[t][1\height]{0.9\columnwidth}%
\includegraphics[clip]{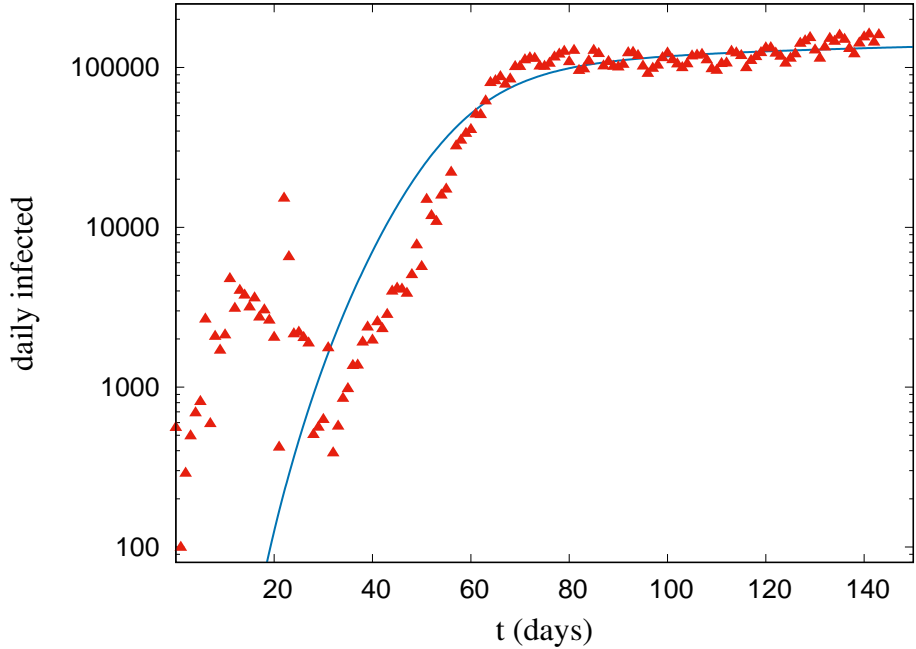}

\caption{Worldwide infections from Jan, 22 to June 13 and fitting with eq.
(\ref{eq:Fitting eq-1-2}). The behavior fits well with parameters
in Table I. Condition (\ref{eq:Condition-1}) is satisfied.}
\end{minipage}
\end{figure}

In the particular case of Mexico the fitting is shown in figure 2.
In this case condition (\ref{eq:Condition-1}) is not fulfilled. In
terms of our model this means that the peak is not predictable within
the present dynamics. Something similar occurs with Brazil, as shown
in figure 3. The data for Brazil neither fulfill the condition (\ref{eq:Condition-1}).
In this case there is neither the prevision of a peak and we can say
that the data for Mexico and Brazil reveals a dynamics where the peak
seems to be quite far if it exists. But there are some illustrative
cases where the peak is reached. Progression of the outbreak in Cuba
and Iceland are shown in Figure 4 and 5 respectively. Condition (\ref{eq:Condition-1})
is satisfied for both countries and we can see that the curve of infection
rate descends at a good speed after past the peak. Now let us take
a look at United States data, shown in Figure 6. The USA outbreak
is characterized by a very fast growth until the peak and, then, very
slow decay of the infection rate is evident. As discussed above, the
outbreak will be controlled for almost infinite time in this dynamics.
There is also some intermediate cases as Spain and Italy, shown in
Figures 7 and 8. In this case the data exhibits the same behavior
as in USA, a fast initial growth and a very slow decay after the peak.
However, the outbreak is controlled in a finite amount of time. In
Table I we present the relevant fitting parameters, including herd
immunity time, $T_{hi}$ and $T_{1000}$, the time to reach a rate
of 1000 infections daily. This, for countries that have not reached
the epidemic peak, Mexico and Brazil. We also include the population,
$P$; of each country. 

\begin{figure}[H]
\begin{minipage}[t]{0.9\columnwidth}%
\includegraphics[clip]{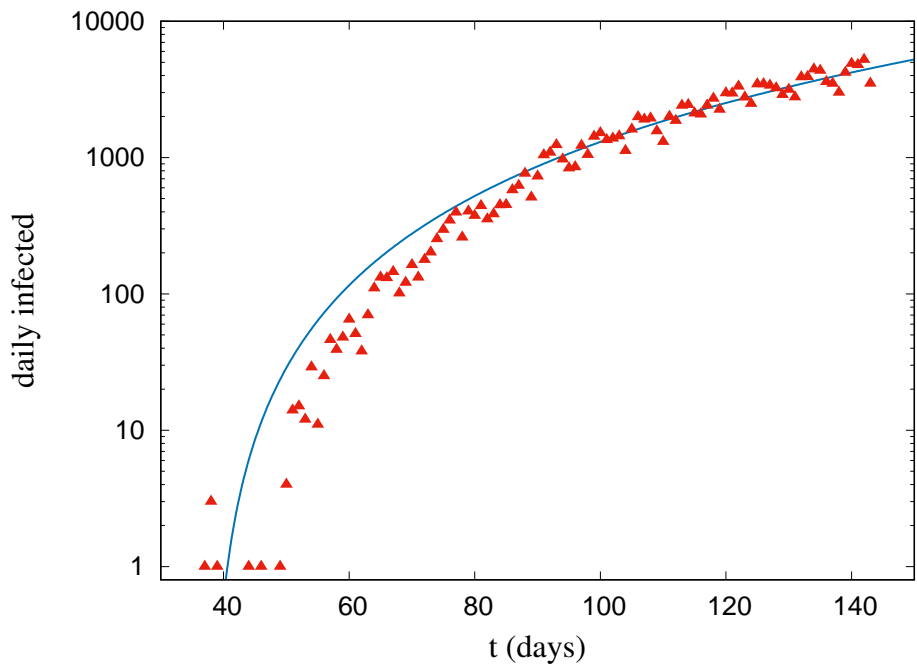}

\caption{Daily infections in Mexico and fitting with eq. (\ref{eq:Fitting eq-1-2})
for parameters in Table I. $T_{hi}=778$ days. Condition (\ref{eq:Condition-1})
is not satisfied.}
\end{minipage}
\end{figure}

\begin{figure}[H]
\begin{minipage}[t]{0.9\columnwidth}%
\includegraphics[clip]{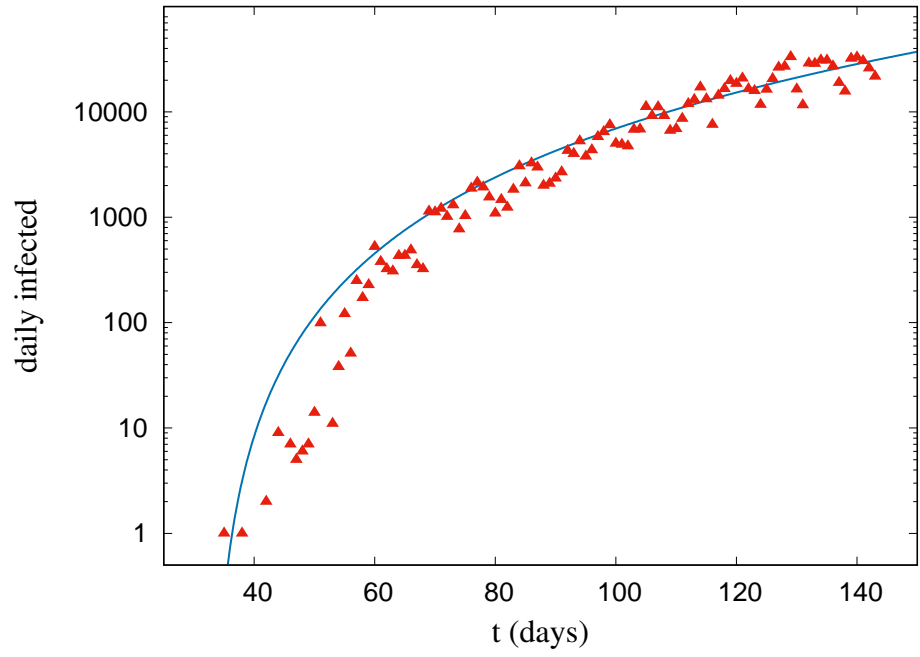}

\caption{Evolution of daily cases in Brazil and fitting with eq. (\ref{eq:Fitting eq-1-2})
for parameters in Table I. $T_{hi}=298$ days. Condition (\ref{eq:Condition-1})
is not satisfied.}
\end{minipage}
\end{figure}

\begin{figure}[H]
\begin{minipage}[t]{0.9\columnwidth}%
\includegraphics[clip]{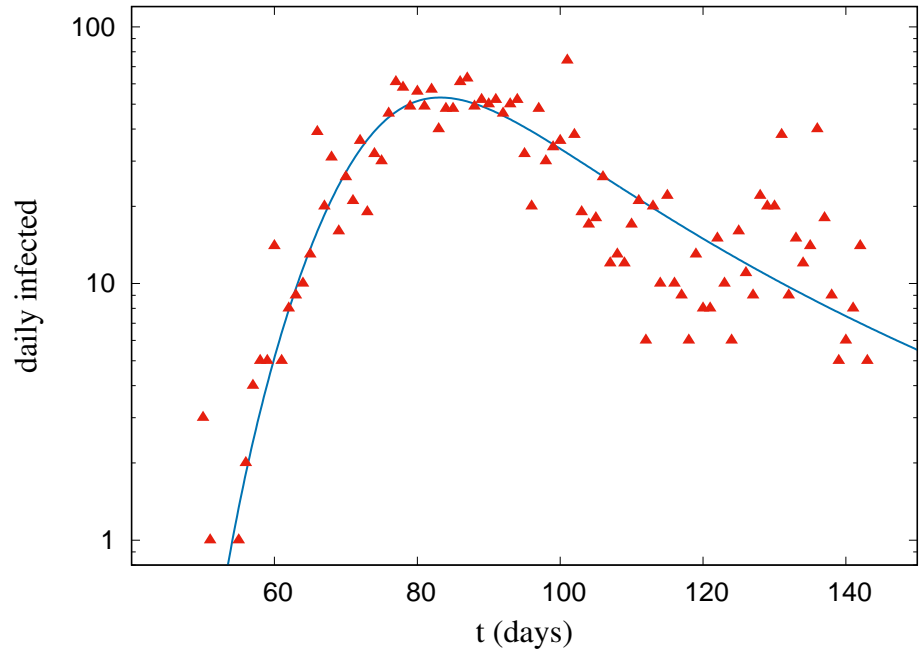}

\caption{Daily infections in Cuba. The theoretical curve fits with data though
with a poor correlation due to the dispersion. See fitting parameters
in Table I. Condition(\ref{eq:Condition-1}) is satisfied.}
\end{minipage}
\end{figure}

\begin{figure}[H]
\begin{minipage}[t]{0.9\columnwidth}%
\includegraphics[clip]{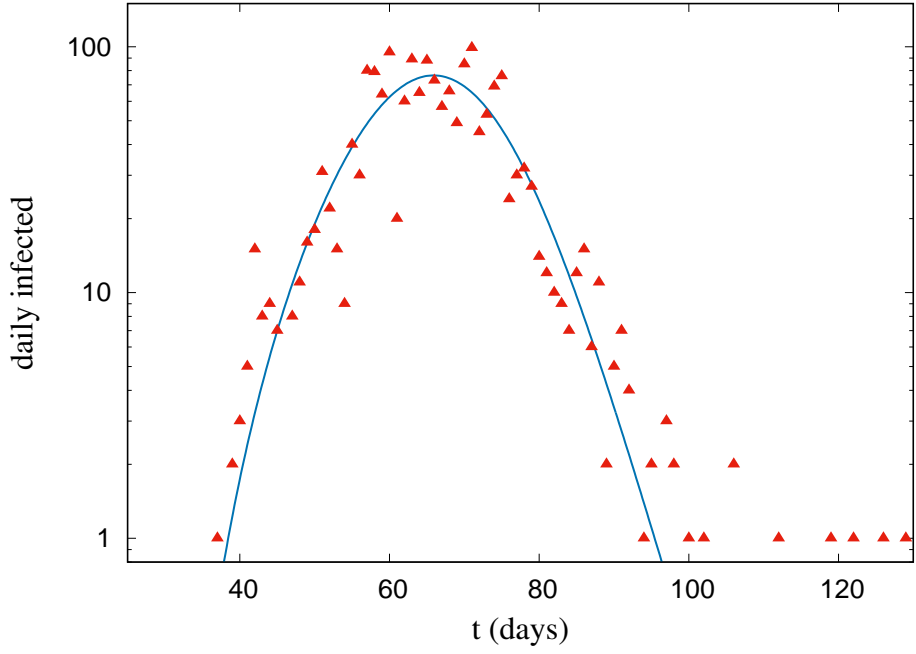}

\caption{Daily infections in Iceland, where the pandemic seems to have ceased.
Here again, in spite of the relatively small correlation coefficient,
the behavior of the pandemic in this country looks well described
by eq. (\ref{eq:Fitting eq-1-2}). See fitting parameters in Table I.
Condition (\ref{eq:Condition-1}) is satisfied.}
\end{minipage}
\end{figure}

\begin{figure}[H]
\noindent\begin{minipage}[t]{1\columnwidth}%
\includegraphics{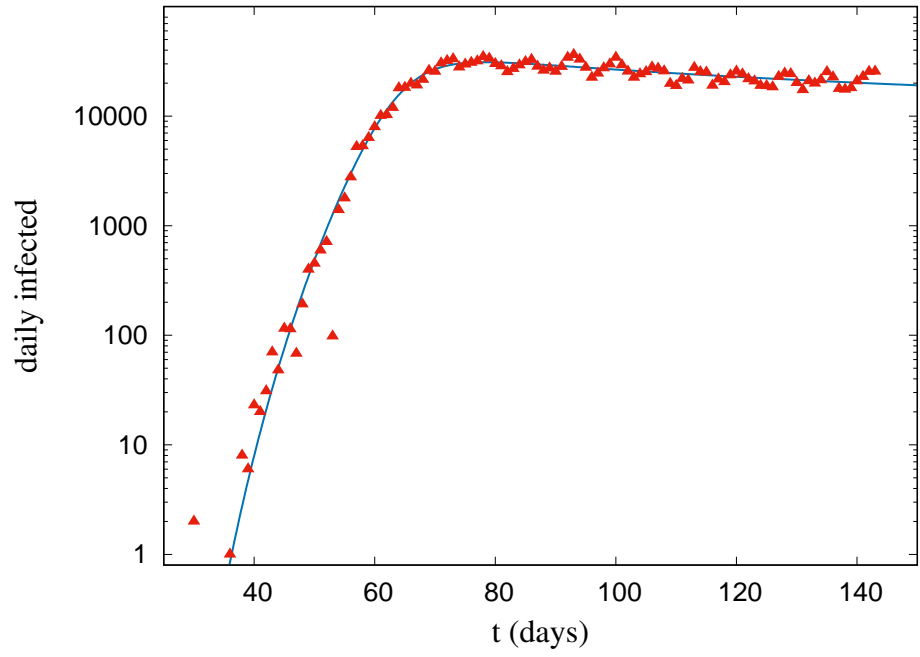}.%
\end{minipage}

\caption{Daily infections in USA, where the peak looks already surpassed. Here
again, the behavior of the pandemic in this country looks well described
by eq. (\ref{eq:Fitting eq-1-2}). See fitting parameters in Table I.
Condition (\ref{eq:Condition-1}) is satisfied.}
\end{figure}

\begin{figure}[H]
\begin{minipage}[t]{0.9\columnwidth}%
\includegraphics[clip]{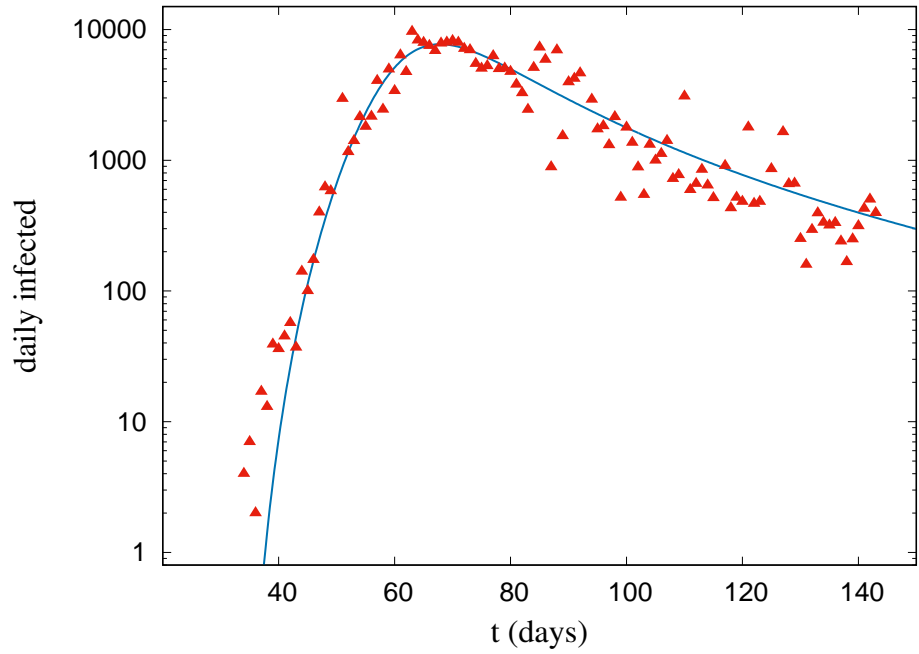}

\caption{Daily infections in Spain. the data shows a large dispersion but the
curve describes well the behavior. See fitting parameters in Table I.
Condition (\ref{eq:Condition-1}) is satisfied.}
\end{minipage}
\end{figure}

\begin{figure}[H]
\begin{minipage}[t]{0.9\columnwidth}%
\includegraphics[clip]{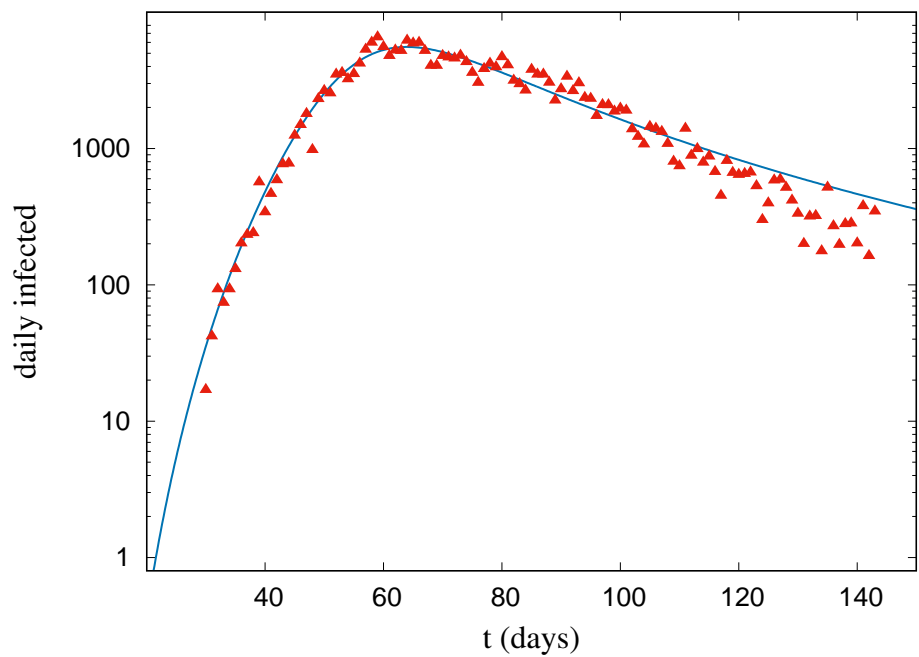}

\caption{Daily infections in Italy. See fitting parameters in Table I. Condition
(\ref{eq:Condition-1}) is satisfied.}
\end{minipage}
\end{figure}

\begin{table}[H]
\begin{minipage}[t]{0.6\columnwidth}%
\label{Table I}%
\begin{tabular}{|c|>{\centering}p{0.2\textwidth}|>{\centering}p{0.2\textwidth}|>{\centering}p{0.2\textwidth}|>{\centering}p{0.2\textwidth}|>{\centering}p{0.2\textwidth}|>{\centering}p{0.2\textwidth}|>{\centering}p{0.2\textwidth}|}
\hline 
Country & A & B & a & b & P & $T_{hi}$(days) & $T_{1000}$\tabularnewline
\hline 
\hline 
Brazil & 0.0152828 & 0.0104434 & 4.31197 & 0.0671095 & 212559417 & 298 & 36\tabularnewline
\hline 
Cuba & 1.80E-05 & 3.30E-09 & 5.31906 & 1.40779 & 11326616 & - & -\tabularnewline
\hline 
Iceland & 6.08E-05 & 1.69E-09 & 5.09845 & 4.94326 & 341243 & - & -\tabularnewline
\hline 
Italy & 1.20E-07 & 1.85E-13 & 7.50956 & 1.32858 & 60461826 & - & 34\tabularnewline
\hline 
Mexico & 0.0541958 & 0.0104956 & 3.60005 & 0.0641971 & 128932753 & 778 & 56\tabularnewline
\hline 
Spain & 0.000170317 & 2.75E-10 & 6.31706 & 1.35476 & 46754778 & - & 19\tabularnewline
\hline 
USA & 1.09E-13 & 5.99E-20 & 11.5099 & 0.973087 & 331002651 & - & 34\tabularnewline
\hline 
Worldwide & 3.18E-06 & 4.65E-13 & 6.84834 & 0.816744 & 7786246434 & - & 29\tabularnewline
\hline 
\end{tabular}%
\end{minipage}

\caption{Relevant fitting parameters for each country.}
\end{table}

As can be seen from fitting coefficients, the exponent $b$ drives
the behavior of infections in every country. Those countries that
manage well the disease expansion have b values wide larger than $1$.
Countries with b values close to one, as Italy and Spain, have managed
the pandemics but poorly and at high costs. The recovery in both countries
will be long. The same is valid for USA, that manage poorly the outbreak
and its struggling with an even longer recovery to normal life. Even
worst scenario is taken place in Mexico and Brazil, with very low
values of $b$. Those countries are experiencing a big outbreak where
even can get herd immunity. This, however, implies very high values
of infections and mortality for the near future. 

But let us briefly comment about herd immunity. Those countries that
have managed to stop the outbreak, even with relative high mortality
as Spain and Italy, will not reach the herd immunity. As a matter
of fact, This can not be calculated for those countries. Then, we
can see countries like Brazil where, if the way of deal with the outbreak
do not change, the herd immunity will be reached. Even when it seems
desirable, the ability to reach the herd immunity brings with it a
high payload. That is, for a country like Brazil the herd immunity
would charge more than 100 million of infected people. That is, much
the same as if a non small war devastates the country. There is an
alike scenario in Mexico, but the difference here is that the value
for $T_{hi}$ is so high that SARS-CoV-2 could even turn into a seasonal
virus, at least for some years. We can expect around the same mortality
but scattered over a few years. 

A special observation deserves USA, where $T_{hi}$ tends to infinity.
Here we can expect a continuous infection rate for a very long time.
The outbreak is controlled but not enough to eradicate the virus.
Virus will not disappear in several years but maybe the healthcare
system could manage it. The virus will get endemic, and immunity will
never be reached. However the infections and mortality rate associated
with it, can be, hypothetically, small if compared with Mexico and
Brazil. We can also compare the speed of the outbreak in different
countries. As we already said in Table I we calculated $T_{1000}$
for some countries. However, it should be noticed that this time is
not calculated from day 0, which is always January 22, but for the
approximated day when the outbreak began in the correspondent country.
By example, in Brazil there was no cases at January, 22 but the first
cases were detected around March, 10. So both, data fitting and $T_{1000}$,
were calculated from March, 10.

\section{Conclusions and Outlook For Further Investigations}

In this work, for the first time, we presented a model built using
the method of analogy, in this case with a nonlinear relaxation-like
behavior. With this, a good fitting with the observed behavior of
the daily number of cases with time is obtained. The explicit expressions
obtained may be used as a tool to approximately forecast the development
of the COVID-19 pandemic in different countries and worldwide. In
principle, this model can be used as a help to elaborate or change
actions. This model does not incorporate any particular property of
this pandemic, so we think it could be used to study pandemics with
different sources. With the collected data of the pandemics at early
times, using this model, it can be predicted the possibility of a
peak, indefinite growth, time for herd immunity, etc.

What seems to be clear from the COVID-19 data, the fitting and the
values shown in the Table I, is that SARS-CoV-2 is far from being
controlled at world level. Even when some countries appear to control
the outbreak, the virus is still a menace for its health system. Furthermore,
in the nowadays interconnected world it is impossible for any country
to keep closed borders and pay attention to what happens only inside.
All isolation measures should be halted at some time and we can expect
new outbreaks in countries like Spain or Italy even after the current
one could be controlled. The only way to control the spread of SARS-CoV-2
seems to be the development of a vaccine that provides the so much
desired herd immunity. Indeed, the model made possible to make an
approximate forecast of the time to reach the herd immunity. This
may be useful in the design of actions and policies about the pandemic.
We have introduced the $T_{1000}$, that gives information about the
early infection behavior in populous countries. A possible improvement
of this model is the formal inclusion of a formulation including the
dual conformable derivative \citep{rosa2018dual,weberszpil2020dual}.
This will be published elsewhere.

\section*{Acknowledgments}

We acknowledge Dr. Carlos Trallero -Giner for helpful comments and
suggestions

\section*{Conflict of Interest }

The authors declare that they have no conflict of interest.

\bibliographystyle{apsrev4-1}
%\addcontentsline{toc}{section}{\refname}\bibliography{reference_covid}
%

\end{document}